\documentclass[aps,showpacs,nofootinbib,superscriptaddress,twocolumn,showkeys,floatfix]{revtex4-1}

\usepackage[italian,english]{babel}
\usepackage[T1]{fontenc}
\usepackage[utf8x]{inputenc}
\usepackage{xcolor}
\usepackage{soul}

\usepackage{graphicx}
\usepackage{epstopdf}         
\graphicspath{{./Images/}}  

\usepackage{booktabs}

\usepackage{amsmath}   
\usepackage{amssymb}   
\usepackage{bm}        
\usepackage{siunitx}   
\usepackage[version=4]{mhchem}    

\usepackage{datetime}

\makeatletter
\preto\maketitle{%
  \begingroup\lccode`~=`,
  \lowercase{\endgroup
  \let\saved@breqn@active@comma~
  \let~}\active@comma 
}
\appto\maketitle{%
  \begingroup\lccode`~=`,
  \lowercase{\endgroup
  \let~}\saved@breqn@active@comma 
}
\makeatother

\begin{document}
\allowdisplaybreaks[2]

\title{Magnetic design of twin aperture $\cos\theta$\\ superconducting dipoles with a semi-analytic approach}

\author{Alessandro Maria Ricci}
\email{alessandro.ricci@ge.infn.it,\\ alessandromaria.ricci@edu.unige.it}
\affiliation{Dipartimento di Fisica, Università di Genova, via Dodecaneso 33, I-16146 Genova, Italy}
\affiliation{INFN sezione di Genova, via Dodecaneso 33, I-16146 Genova, Italy}

\author{Pasquale Fabbricatore}
\email{pasquale.fabbricatore@ge.infn.it}
\affiliation{INFN sezione di Genova, via Dodecaneso 33, I-16146 Genova, Italy}

\author{Stefania Farinon}
\email{stefania.farinon@ge.infn.it}
\affiliation{INFN sezione di Genova, via Dodecaneso 33, I-16146 Genova, Italy}

\begin{abstract}
The magnetic design is a basic aspect of the superconducting magnets for particle accelerators. When dealing with single aperture $\cos\theta$ type dipoles, at the first order, the design can be performed with an analytic approach based on a sector dipole approximation followed by a numerical optimization.  For double aperture dipoles the magnetic cross-talk between apertures makes this approach unfeasible. We have developed a semi-analytic model, which starting from a sector dipole approximation, allows to consider the cross-talk between the two apertures. We also demonstrate that the iron yoke contribution to harmonics, although dominant, does not change the optimal configuration found in its absence. As examples, we show two possible electromagnetic designs for the D2 dipole of the High Luminosity upgrade of LHC. The semi-analytic model can be generalized to a larger class of magnets.
\end{abstract}

\date{\today, \currenttime}


\maketitle

\section{Introduction}
The superconducting dipoles bending the beams in particle accelerators must provide  a high homogeneous  magnetic field. The generally used criterion is that any higher order multipole component must be lower than $10^{-4}$ of the central field. Moreover, many constraints on the coil shape (minimum bending radius, maximum magnet dimensions, inter-layer spacers, ...), the operating margins, the effects of permanent currents and of magnetic components and the costs have to be taken into account, introducing difficulties in the design.

Four different types of layouts have been built and tested over the years to generate magnetic dipoles: $\cos\theta$ coil~\cite{CERN:LHC}, common coil~\cite{Novitski:common coil}, block coil~\cite{Milanese:FRESCA2} and more recently canted $\cos\theta$ solenoid~\cite{Caspi:CCT}. The most used configuration is the $\cos\theta$ type, which can be considered a simple way for approximating an ideal annular current density distribution proportional to the cosine of the azimuth, so generating a perfect dipole. In practical layouts, the annulus is approximated by conveniently piling up the conductors in blocks separated by spacers and carrying the same constant current density. This arrangement, with different number of layers and of spacers, has been widely used for the dipoles built until now~\cite{CERN:LHC, Wilson:Tevatron}. Presently, most of the magnets for the High Luminosity upgrade~\cite{CERN:HL-LHC} of the Large Hadron Collider~\cite{CERN:LHC} at CERN are based on this layout and EuroCirCol Collaboration lately has chosen the $\cos\theta$ design as baseline for the dipoles of the Future Circular Collider~\cite{Schoerling:EuroCirCol}.

The dipolar coils typically include a long straight section ($>1$~\si{m}), so that a 2D analysis assuming infinitely long conductors can be considered a good approximation. For the $\cos\theta$ layout, many numerical algorithms exist to optimize the position of the conductors in the cross section~\cite{Russenschuck:field computation}, but all of them, to be really effective, have to operate on configurations which are not too far from a local optimum. Analytic models of $\cos\theta$ coils have been done for dipoles and quadrupoles~\cite{Devred:sc magnets}, approximating the blocks as annular sectors, and are often used to carry out an initial coarse optimization of the parameters of an accelerator and to estimate dimensions and costs~\cite{Rossi:sector coil}. However, they can't be employed to control the homogeneity of the magnetic field for various reasons. First of all, in a real coil the block shape differs from the sector. Moreover, in colliders as LHC, where two beams run one very near the other, the magnets must be done in twin aperture, i.e. with two close coils surrounded by the iron yoke. The sector model can describe analytically neither the cross-talk between the two coils nor the non-linear iron yoke contribution. This problem is particularly important to a special class of dipoles involved in proximity of the Interaction Region (IR) of colliders, the separation/recombination dipoles. These special magnets are used for separating and recombining the beams before and after the collision in the Interaction Point (IP). In order to accomplish this role the magnetic field must be concord in both apertures generating a considerable magnetic cross-talk if the magnetic field is enough high as it happens for the D2 magnets of the High Luminosity upgrade of LHC~\cite{CERN:HL-LHC, Farinon:D2}.

Starting from the analytic model of a sector dipole, we show a semi-analytic model that lets to control and optimize the field quality of $\cos\theta$ dipoles in twin aperture, supposing that the iron yoke contribution to harmonics, although dominant, does not change the optimal configuration found in its absence. It is worth noting that the same approach can be used for different coil layouts as for instance dipoles in block coil and common coil configurations, as reported in Ref.~\cite{Rochepault:block coil, Xu:common coil}, or quadrupoles.

In Section II we review the sector model, in Section III we introduce the semi-analytical model and finally in Section IV we show as examples two possible electromagnetic designs for the D2 dipole~\cite{Farinon:D2} of the High Luminosity upgrade of the Large Hadron Collider.

\section{The sector model}

It is well known that for the accelerator magnets, away from the ends, the multipole fields have only components on $xy$-plane and they are constant along the beam (i.e. along the magnet). So, the vector potential has only the component $A_z$ and we can resolve the Laplace equation in cylindrical coordinates. Then, introducing the complex notation, the magnetic field can be written as:
\begin{equation}
B=B_y+iB_x=\sum_{n=1}^\infty\left(B_n+iA_n\right)\left(\frac{x+iy}{R_{ref}}\right)^{n-1} \,,
\label{eq:magnetic field}
\end{equation}
where $R_{ref}$ is a reference radius usually chosen as $2/3$ of the aperture radius and the coefficients $A_n$ and $B_n$ are called skew and normal cylindrical harmonics, respectively. In European definition~\eqref{eq:magnetic field}, each component of order $n$ represents the $2n$-pole component. In polar coordinates the cylindrical harmonics are expressed as
\begin{align}
&A_n=\frac{n}{\pi R_{ref}}\int_0^{2\pi}A_z\left(R_{ref},\theta\right)\sin n\theta\, d\theta \,, \\[0.3cm]
&B_n=-\frac{n}{\pi R_{ref}}\int_0^{2\pi}A_z\left(R_{ref},\theta\right)\cos n\theta\, d\theta \,,
\end{align}
where $A_z(R_{ref},\theta)$ is the vector potential calculated at the reference radius. 

The cylindrical harmonics of a dipole can be normalized to units as $b_n=10^{4}B_n/B_1$ and $a_n=10^{4}A_n/B_1$, where $B_1$ is the dipole field, so that eq.~\eqref{eq:magnetic field} becomes
\begin{equation}
B_y+iB_x=10^{-4}B_1\sum_{n=1}^\infty\left(b_n+ia_n\right)\left(\frac{x+iy}{R_{ref}}\right)^{n-1} \,.
\end{equation}

Integrating the vector potential $A_z=-\frac{\mu_{0}I}{2\pi}ln\frac{r}{\rho}$ generated by a current line $I$ in position $(\rho,\phi)$, where $r^{2}=\rho^{2}+R_{ref}^{2}-2\rho R_{ref}cos\left(\phi-\theta\right)$, we find the normal multipole coefficients for a current line with $R_{ref}<\rho$:
\begin{equation}
B_n\left(\rho,\phi\right)=-\frac{\mu_0 I}{2\pi R_{ref}}\left(\frac{R_{ref}}{\rho}\right)^n \cos n\phi \,.
\label{eq:harmonics of a current line}
\end{equation}
Similarly, we can find the skew coefficients $A_n(\rho,\phi)$.

Integrating this last equation over the regions on the xy-plane where $I\neq 0$, one can find analytic or semi-analytic expressions for the harmonic components generated by any system of currents.
\begin{figure}[htbp]
\includegraphics[width=0.4\textwidth]{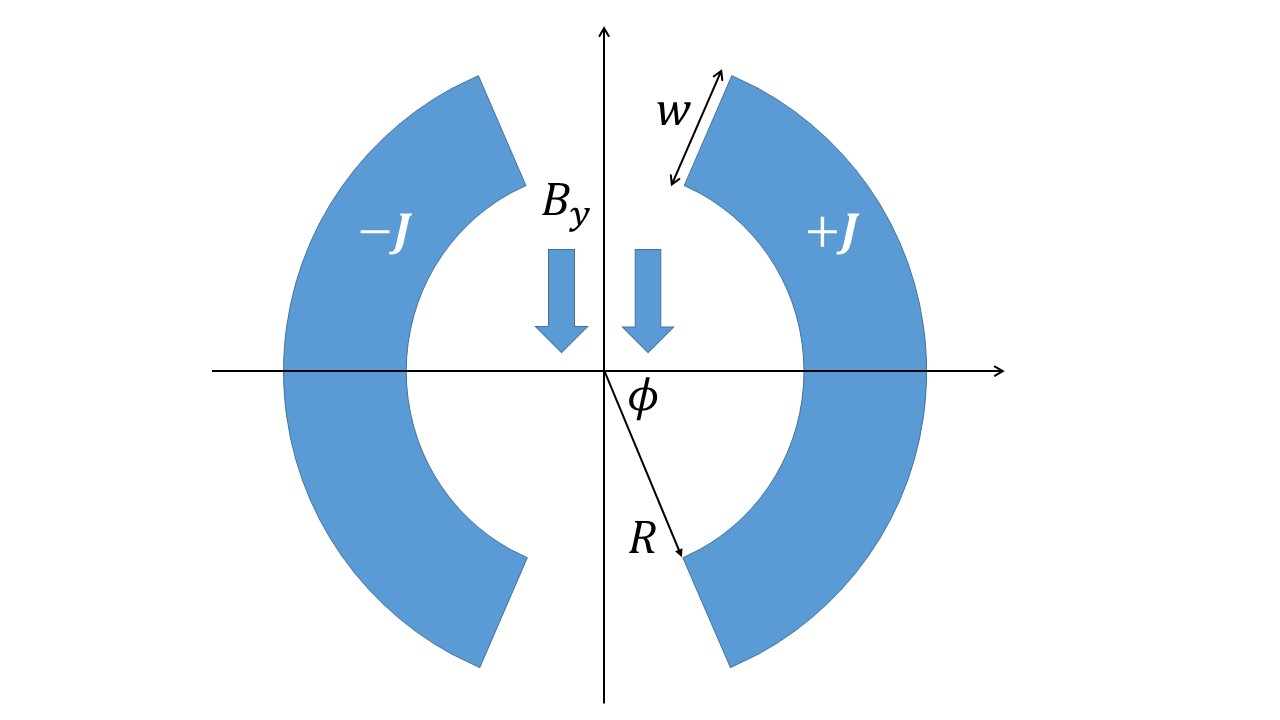}
\caption{Sector coil layout for a dipole of inner radius $R$ and coil width $w$, spanning the angle from $-\phi$ to $\phi$. In the right half coil an uniform current density $J$ flows, in the left $-J$. The magnetic field in the middle has the component $B_y$ only.}
\label{fig:sector dipole}
\end{figure}

For instance, if we consider a single block dipole as shown in Fig.~\ref{fig:sector dipole}, with a uniform current density $J$, the odd normal harmonics can be obtained by integrating eq.~\eqref{eq:harmonics of a current line} over the sector:
\begin{equation}
B_n=-\frac{2\mu_0 J R_{ref}^{n-1}}{\pi n\left(n-2\right)}\left(\frac{1}{R^{n-2}}-\frac{1}{\left(R+w\right)^{n-2}}\right)\sin n\phi \,.
\label{eq:harmonics for sector}
\end{equation}

In case of left-right asymmetric coil, which has to be introduced to minimize the effects of cross-talk in twin aperture magnets, as will be shown in the next section, the normal multipoles becomes:
\begin{equation}
\begin{split}
B_n=-&\frac{\mu_0 J R_{ref}^{n-1}}{\pi n\left(n-2\right)}\left(\frac{1}{R^{n-2}}-\frac{1}{\left(R+w\right)^{n-2}}\right) \\[0.3cm]
&\times\left(\sin n\phi-\left(-1\right)^n \sin n\psi\right) \quad n\neq 2
\end{split}
\label{eq:asymmetric configuration}
\end{equation}
and
\begin{equation}
B_2=-\frac{\mu_0 J R_{ref}}{2\pi}\ln{\frac{R+w}{R}}\left(\sin 2\phi-\sin 2\psi\right) \,,
\label{eq:asymmetric configuration 2}
\end{equation}
where $\phi$ is the angle for the right sector and $\psi$ is the angle for the left sector.

\section{The semi-analytic model}

The twin aperture configuration introduces a complicating factor, i.e. the evaluation of the contribution to the harmonic components which one aperture exerts on the other. For this reason, we propose a semi-analytic model, which extend the sector model and is based on three statements.

\begin{enumerate}
\item We suppose that the iron yoke contribution to harmonics, although dominant, does not change the optimal configuration found in its absence.
\item We assume that the difference between sectors and real blocks is small, so we can describe analytically one coil (e.g. the right one) using a discretized sector model. Each sector is identified by a starting angle $\phi_{i}$ and by a number of turns $m_{i}$, where $i$ is an index identifying the sector number. The total angle spanned by each sector is given by $m_i d \phi$, where $d\phi$ is the ``quantum'' of angle occupied by each turn. It is calculated as
\begin{equation}
d\phi=\arcsin\frac{l}{R} \,,
\end{equation}
where $l$ is the middle thickness of the cable considered as conductor plus insulation and $R$ is the inner radius.
\item Because the cables of the left coil are far from the center of the right coil (i.e. from the region where harmonics are computed), we can describe analytically also the left coil, approximating each turn with a single current line flowing in the center of the turn itself.
\end{enumerate}

\begin{figure}[htbp]
\includegraphics[width=0.4\textwidth]{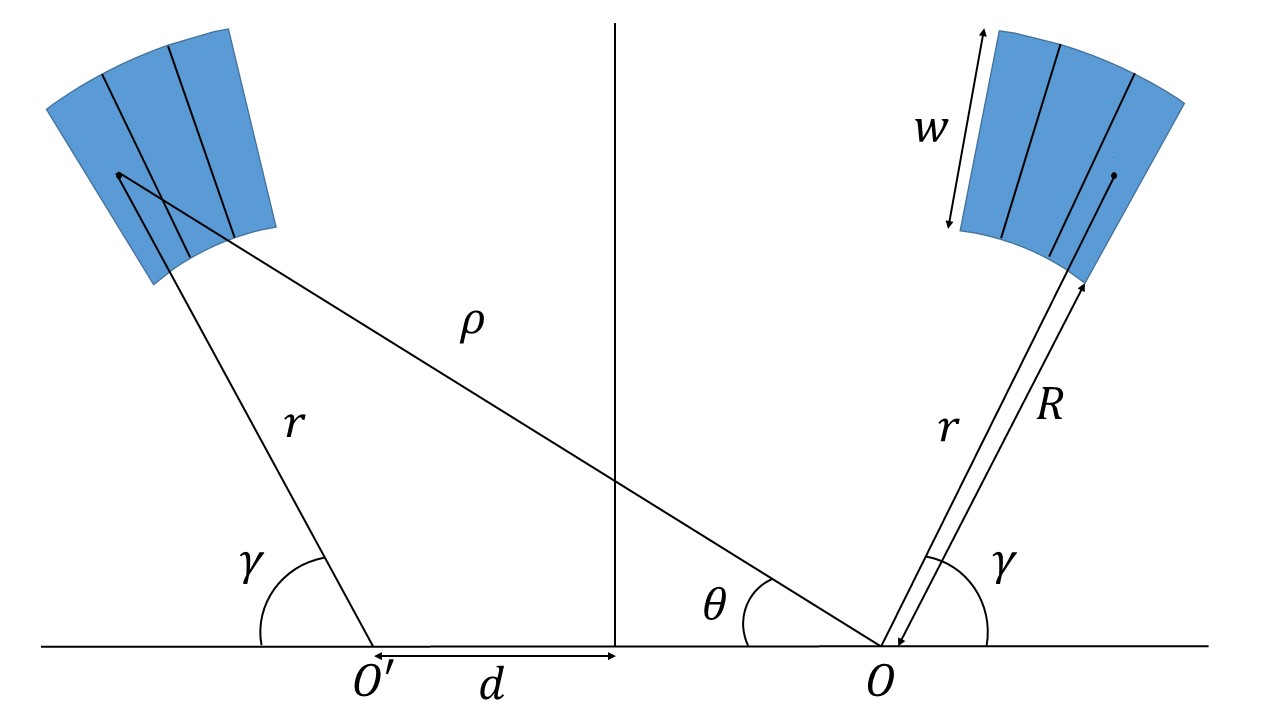}
\caption{Polar coordinates of the current lines for the external sectors of the two coils. Only one sector is shown in the plot.}
\label{fig:specular blocks}
\end{figure}

\begin{figure}[htbp]
\includegraphics[width=0.4\textwidth]{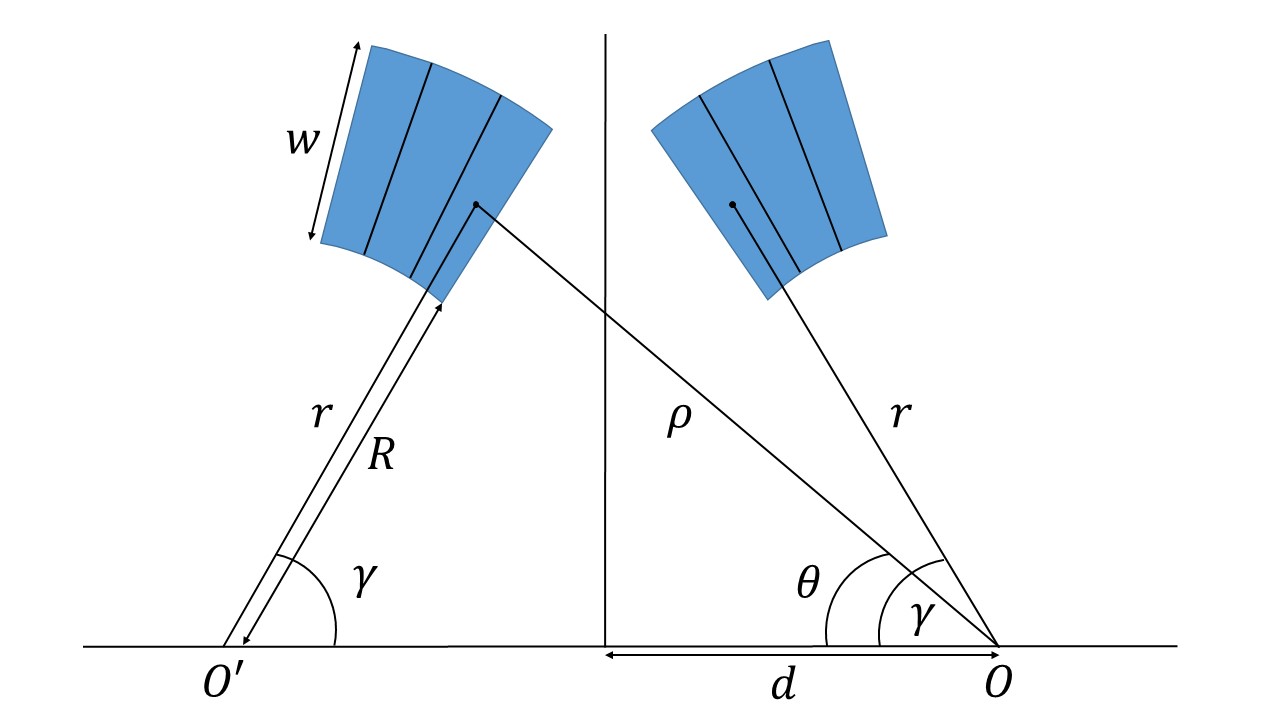}
\caption{Polar coordinates of the current lines for the internal sectors of the two coils. Only one sector is shown in the plot.}
\label{fig:specular blocks 2}
\end{figure}

Because the left coil is mirrored to the right one, we must connect the coordinates of the current lines $(\rho_{ij},\theta_{ij})$, where $j$ is an integer from $0$ to $m_i-1$, to the variables of each sector $i$ $(\phi_{i},m_{i})$. This can be done by simple trigonometric formulas. First, we define the polar coordinates of the current lines in the middle of each turn of the right coil (see Fig.~\ref{fig:specular blocks}-\ref{fig:specular blocks 2}) as
\begin{equation}
\begin{aligned}
&r=R+\frac{w}{2} \,, \\[0.3cm]
&\gamma_{ij}=\phi_i+\left(j+\frac{1}{2}\right)d\phi \,.
\end{aligned}
\label{eq:trigonometry}
\end{equation}
Then, we set the polar coordinates of the current lines of the left coil, splitting between external and internal sectors of each coil (see always Fig.~\ref{fig:specular blocks}-\ref{fig:specular blocks 2}). For the external sectors we obtain
\begin{equation}
\begin{aligned}
&\theta_{ij}=\arctan\left(\frac{r\sin\gamma_{ij}}{2d+r\cos\gamma_{ij}}\right) \,, \\[0.3cm]
&\rho_{ij}=\frac{2d+r\cos\gamma_{ij}}{\cos\theta_{ij}} \,,
\end{aligned}
\label{eq:external trigonometry}
\end{equation}
where $d$ is half of the inter-beam distance; while for the internal sectors we find
\begin{equation}
\begin{aligned}
&\theta_{ij}=\arctan\left(\frac{r\sin\gamma_{ij}}{2d-r\cos\gamma_{ij}}\right) \,, \\[0.3cm]
&\rho_{ij}=\frac{2d-r\cos\gamma_{ij}}{\cos\theta_{ij}} \,.
\end{aligned}
\label{eq:internal trigonometry}
\end{equation}


The algorithm is performed in the following way. First of all, we create a random symmetric configuration $\left(\phi_1, m_1, \phi_2, m_2, \dots, \phi_N, m_N\right)$ of the right coil, solving numerically for each $n$ odd the equation system for a sector model with fixed number of blocks $N$:
\begin{equation}
\begin{split}
    b_{n}&\left(\phi_1, m_1, \dots, \phi_N, m_N\right) \\[0.3cm]
    &=10^4\frac{B_n\left(\phi_1, m_1, \dots, \phi_N, m_N\right)}{B_1\left(\phi_1, m_1, \dots, \phi_N, m_N\right)}=0 \, ,
\end{split}
\label{eq:equation system}
\end{equation}
where $B_1$ and $B_n$ are given by eq.~\eqref{eq:harmonics for sector}:
\begin{align}
\begin{split}
    &B_1\left(\phi_1, m_1, \dots, \phi_N, m_N\right) \\[0.3cm]
    &\qquad=-\frac{2\mu_0 J w}{\pi} \sum_{i=1}^N \bigl[\sin{\left(\phi_i+m_i d\phi\right)}-\sin{\phi_i}\bigr].
\end{split}
\\[0.3cm]
\begin{split}
    &B_n\left(\phi_1, m_1, \dots, \phi_N, m_N\right) \\[0.3cm]
    &\qquad=-\frac{2\mu_0 J R_{ref}^{n-1}}{\pi n \left(n-2\right)}\left(\frac{1}{R^{n-2}}-\frac{1}{\left(R+w\right)^{n-2}}\right) \\[0.3cm]
    &\qquad\qquad\qquad\times \sum_{i=1}^N \bigl[\sin{n\left(\phi_i+m_id\phi\right)}-\sin{n\phi_i}\bigr],
\end{split}
\end{align}
Then, we mirror this configuration to the left side, i.e. we compute the coordinates of the current lines of the left coil, using eq.~\eqref{eq:trigonometry},~\eqref{eq:external trigonometry} and~\eqref{eq:internal trigonometry}, and the sum of their contributions to the harmonic components of the right aperture, by eq.~\eqref{eq:harmonics of a current line} with fixed current intensity $I$. For the left sectors of left coil we use the equation
\begin{equation}
\begin{split}
  B_n&\left(\rho_{10}^{ll}, \theta_{10}^{ll}, \rho_{11}^{ll},\theta_{11}^{ll}, \dots, \rho_{ij}^{ll},\theta_{ij}^{ll}, \dots, \rho_{N\,m_N-1}^{ll},\theta_{N\,m_N-1}^{ll}\right) \\[0.3cm]
  &=-\frac{\mu_0 \left(-I\right) R_{ref}^{n-1}}{\pi}\sum_{i=1}^N\sum_{j=0}^{m_i-1}\frac{\cos n\left(\pi-\theta_{ij}^{ll}\right)}{\left(\rho_{ij}^{ll}\right)^n} \\[0.3cm]
  &=-(-1)^n\frac{\mu_0 \left(-I\right) R_{ref}^{n-1}}{\pi}\sum_{i=1}^N\sum_{j=0}^{m_i-1}\frac{\cos n\theta_{ij}^{ll}}{\left(\rho_{ij}^{ll}\right)^n} \,,
\end{split}
\label{eq:left-left coil harmonics}
\end{equation}
where $\rho_{ij}^{ll}$ and $\theta_{ij}^{ll}$ are given by eq.~\eqref{eq:external trigonometry}. Likewise, for the right sectors we use the equation
\begin{equation}
\begin{split}
  B_n&\left(\rho_{10}^{lr}, \theta_{10}^{lr}, \rho_{11}^{lr},\theta_{11}^{lr}, \dots, \rho_{ij}^{lr},\theta_{ij}^{lr}, \dots, \rho_{N\,m_N-1}^{lr},\theta_{N\,m_N-1}^{lr}\right) \\[0.3cm]
  &=-\frac{\mu_0 I R_{ref}^{n-1}}{\pi}\sum_{i=1}^N\sum_{j=0}^{m_i-1}\frac{\cos n\left(\pi-\theta_{ij}^{lr}\right)}{\left(\rho_{ij}^{lr}\right)^n} \\[0.3cm]
  &=-(-1)^n\frac{\mu_0 I R_{ref}^{n-1}}{\pi}\sum_{i=1}^N\sum_{j=0}^{m_i-1}\frac{\cos n\theta_{ij}^{lr}}{\left(\rho_{ij}^{lr}\right)^n} \,,
\end{split}
\label{eq:left-right coil harmonics}
\end{equation}
where $\rho_{ij}^{lr}$ and $\theta_{ij}^{lr}$ are given by eq.~\eqref{eq:internal trigonometry}. So, the $n$ order contribution of the left coil is
\begin{equation}
\begin{split}
    k_n&=B_n\left(\rho_{10}^{ll}, \theta_{10}^{ll}, \dots, \rho_{ij}^{ll},\theta_{ij}^{ll}, \dots, \rho_{N\,m_N-1}^{ll},\theta_{N\,m_N-1}^{ll}\right) \\[0.3cm]
    & + B_n\left(\rho_{10}^{lr}, \theta_{10}^{lr}, \dots, \rho_{ij}^{lr},\theta_{ij}^{lr}, \dots, \rho_{N\,m_N-1}^{lr},\theta_{N\,m_N-1}^{lr}\right) \,.
    \label{eq:left coil contribution}
\end{split}
\end{equation}
Finally, we resolve numerically a new equation system to find a new coordinate set of an asymmetric configuration $\left(\phi_1^{'}, \psi_1, m_1^{'}, \phi_2^{'}, \psi_2, m_2^{'}, \dots, \phi_N^{'}, \psi_N, m_N^{'}\right)$ which offsets the harmonics~\eqref{eq:left coil contribution} regarding them as fixed values:
\begin{equation}
\begin{split}
    &b_n\left(\phi_1^{'}, \psi_1, m_1^{'}, \dots, \phi_N^{'}, \psi_N, m_N^{'}\right) \\[0.3cm]
    &\quad=10^4\frac{B_n\left(\phi_1^{'}, \psi_1, m_1^{'}, \dots,\phi_N^{'}, \psi_N, m_N^{'}\right)+k_n}{B_1\left(\phi_1^{'}, \psi_1, m_1^{'}, \dots, \phi_N^{'}, \psi_N, m_N^{'}\right)+k_1}=0 \,,
\end{split}
\label{eq:equation system 2}
\end{equation}
where $B_1$, $B_2$ and $B_n$ are got from eq.~\eqref{eq:asymmetric configuration} and~\eqref{eq:asymmetric configuration 2}:
\begin{align}
    \begin{split}
        B_1&\left(\phi_1^{'}, \psi_1, m_1^{'}, \dots,\phi_N^{'}, \psi_N, m_N^{'}\right)  \\[0.3cm]
        &=-\frac{\mu_0 J w}{\pi} \sum_{i=1}^N \biggl[\sin \left(\phi_i^{'}+m_i^{'}\,d\phi\right)-\sin \phi_i^{'}  \\[0.3cm]
        &\quad\qquad\qquad\qquad+\sin\left(\psi_i+m_i^{'}\,d\phi\right)-\sin\psi_i\biggr] \,,
    \end{split}
\\[0.5cm]
    \begin{split}
        B_2&\left(\phi_1^{'}, \psi_1, m_1^{'}, \dots,\phi_N^{'}, \psi_N, m_N^{'}\right)  \\[0.3cm]
        &=-\frac{\mu_0 J R_{ref}}{2\pi}\ln{\frac{R+w}{R}}  \\[0.3cm]
        &\qquad\times\sum_{i=1}^N \biggl[\sin 2\left(\phi_i^{'}+m_i^{'}\,d\phi\right)-\sin 2\phi_i^{'}  \\[0.3cm]
        &\quad\qquad\qquad-\sin 2\left(\psi_i+m_i^{'}\,d\phi\right)+\sin 2\psi_i\biggr] \,,
    \end{split}
\\[0.5cm]
    \begin{split}
        B_n&\left(\phi_1^{'}, \psi_1, m_1^{'}, \dots,\phi_N^{'}, \psi_N, m_N^{'}\right)  \\[0.3cm]
        &=-\frac{\mu_0 J R_{ref}^{n-1}}{\pi n \left(n-2\right)}\left(\frac{1}{R^{n-2}}-\frac{1}{\left(R+w\right)^{n-2}}\right)  \\[0.3cm]
        &\;\times \sum_{i=1}^N \biggl[\sin n\left(\phi_i^{'}+m_i^{'}\,d\phi\right)-\sin n\phi_i^{'}\,  \\[0.3cm]
        &\;-\left(-1\right)^n\left(\sin n\left(\psi_i+m_i^{'}\,d\phi\right)-\sin n\psi_i\right)\biggr] \quad \forall n\geq 3 \,,
    \end{split}
\end{align}
with the current density of each conductor in the right coil blocks $J$ derived from the current intensity $I$, as $J=I/S$, where $S$ is the area of each conductor computed as
\begin{equation}
    S=\frac{\left(R+w\right)^2-R^2}{2}\sin{d\phi}=\frac{l\left[\left(R+w\right)^2-R^2\right]}{2R} \,.
\end{equation}

 The equation systems~\eqref{eq:equation system} and~\eqref{eq:equation system 2} have $2N$ and $3N$ freedom degrees, respectively. So, eq.~\eqref{eq:equation system} can be solved to set to zero the first $2N$ harmonics and eq.~\eqref{eq:equation system 2} for the first $3N$ harmonics. We can put additional constraints to rule out unrealistic configurations and set the total number of turns in the coil. Then we proceed by mirroring again this new configuration to the left side to the harmonic components of the right aperture and then resolving again the equation system~\eqref{eq:equation system 2}. We repeat these steps until the configuration doesn't change anymore. Finally, we use this result as a starting point for a numerical optimization which considers the real shape of the blocks and the iron yoke contribution.

\section{Numerical results}

The High Luminosity upgrade~\cite{CERN:HL-LHC} of the Large Hadron Collider~\cite{CERN:LHC} at CERN requires the replacement of the superconducting magnets before and after the interaction regions (IRs) of the ATLAS and CMS experiments~\cite{Bottura:magnets}. An important role is played by the dipoles recombining and separating the particles of the two proton beams around the Interaction Regions (IRs)~\cite{Todesco:IR}. This section is made up of two dipoles, D1 and D2, which bend the two beams in opposite directions. In particular D2~\cite{Farinon:D2} is a twin aperture magnet (both apertures are $105$~\si{\milli\meter} in diameter) with an interbeam distance of $188$~\si{\milli\meter}, generating in both apertures an integrated dipolar magnetic field of $35$~\si{\tesla\meter} with the same polarity. The coil is wound with the same conductor as the LHC dipole outer layer~\cite{CERN:LHC}.
\begin{table}[htbp]
\caption{Main features of D2 dipole}
\begin{tabular}{c c c}
\toprule
\toprule
Feature & Unit & Value \\
\midrule
Bore magnetic field & \si{\tesla} & $4.5$ \\
Magnetic length & \si{\meter} & $7.78$ \\
Peak field & \si{\tesla} & $5.26$ \\
Operating current & \si{\kilo\ampere} & $12.34$ \\
Stored energy & \si{\mega\joule} & $2.28$ \\
Overall current density & \si{\ampere\per\square\milli\meter} & $443$ \\
Aperture & \si{\milli\meter} & $105$ \\
Separation beam at cold & \si{\milli\meter} & $188$ \\
Operating temperature & \si{\kelvin} & $1.9$ \\
Margin on load line & \% & $33$ \\
Multipole variation due to iron saturation & unit & $<10$ \\
\bottomrule
\bottomrule
\end{tabular}
\label{tab:D2}
\end{table}
\begin{figure}[htbp]
\includegraphics[width=0.3\textwidth]{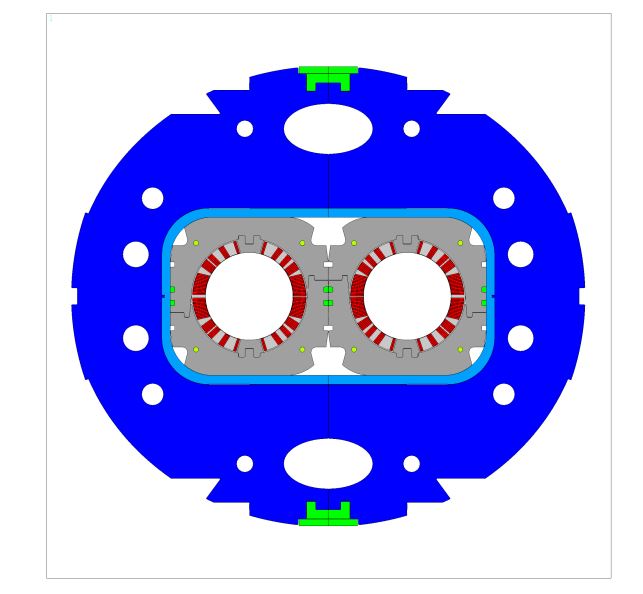}
\caption{Schematic view of the D2 cold mass. Main components are the conductors, in red, the copper wedges, in light grey, the stainless steel collars, in grey, the \ce{AI} alloy sleeves, in light blue, the iron yoke, in blue, and stainless steel keys, pins and clamps in green.}
\label{fig:D2}
\end{figure}
The main features of the D2 dipole are listed in Table~\ref{tab:D2} and Fig.~\ref{fig:D2} shows a schematic view of the cold mass. The main components are the winding (in red) split into five blocks for a total of 31 turns per quadrant, the copper wedges (in light grey), the stainless steel collars (in grey), the \ce{Al} alloy sleeves (in light blue), the iron yoke (in blue) and stainless steel keys, pins and clamps (in green). Each aperture is individually collared, then both are inserted into the \ce{Al} alloy sleeves, whose function is keeping the apertures in the right position and support the repulsive Lorentz force, nearly $0.2$~\si{\mega\newton\per\meter}, arising at full current. The cross-talk between the two coils is compensated through a left-right asymmetric coil design.

This dipole was designed in last years at INFN and a short model has been constructed~\cite{Bersani:D2} and presently under test. The magnetic design was performed with Roxie starting from a tentative initial configuration based on some analytical considerations~\cite{Rossi:sector coil}. We have reconsidered this design on the basis of the developed semi-analytical approach and studied the configurations with three, four and five asymmetric blocks.
\begin{figure}[htbp]
\includegraphics[width=0.45\textwidth]{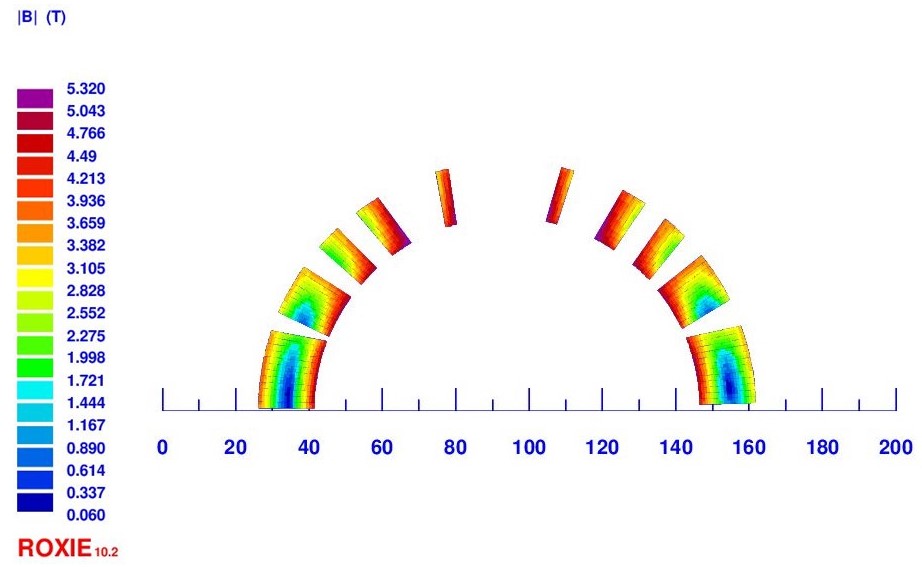}
\caption{Asymmetric configuration with 5 blocks.}
\label{fig:D2 5 blocks}
\end{figure}
\begin{table}[htbp]
\caption{Normal harmonics at operating current for the configuration with $5$ blocks.}
\begin{tabular}{c c c c c c c c c c c c}
\toprule
\toprule
$b_{2}$ & $b_{3}$ & $b_{4}$ & $b_{5}$ & $b_{6}$ & $b_{7}$ & $b_{8}$ & $b_{9}$ & $b_{10}$ & $b_{11}$ & $b_{12}$ & $b_{13}$ \\
\midrule
$-0.02$ & $0$ & $-0.01$ & $0$ & $0$ & $0.04$ & $0$ & $0$ & $0.80$ & $-0.74$ & $-0.21$ & $0.65$ \\
\bottomrule
\bottomrule
\end{tabular}
\label{tab:harmonics 5 blocks}
\end{table}
\begin{figure}[htbp]
\includegraphics[width=0.45\textwidth]{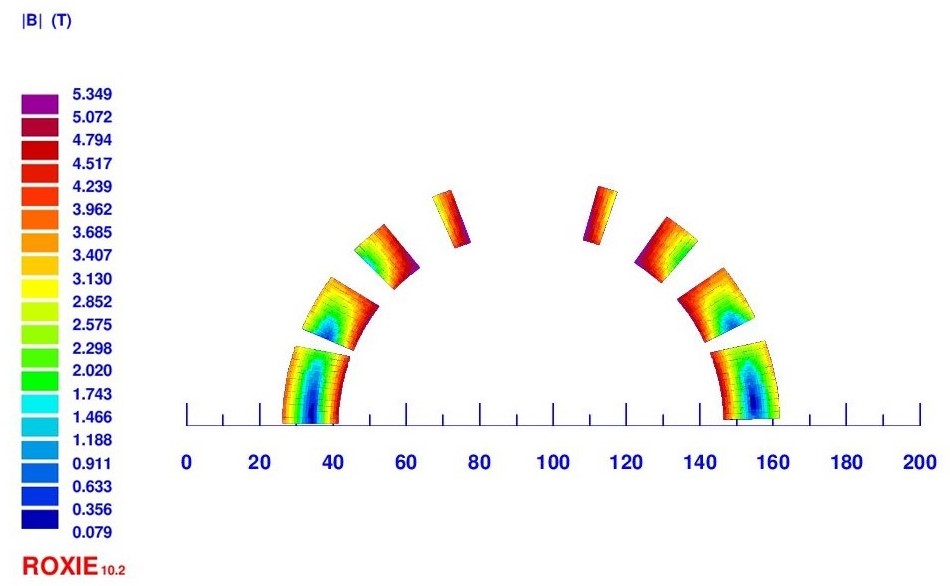}
\caption{Asymmetric configuration with 4 blocks.}
\label{fig:D2 4 blocks}
\end{figure}
\begin{table}[htbp]
\caption{Normal harmonics at operating current for the configuration with $4$ blocks.}
\begin{tabular}{c c c c c c c c c c c c}
\toprule
\toprule
$b_{2}$ & $b_{3}$ & $b_{4}$ & $b_{5}$ & $b_{6}$ & $b_{7}$ & $b_{8}$ & $b_{9}$ & $b_{10}$ & $b_{11}$ & $b_{12}$ & $b_{13}$ \\
\midrule
$0$ & $0.02$ & $0$ & $0$ & $0$ & $-0.55$ & $0$ & $0.21$ & $-0.32$ & $-0.09$ & $-0.34$ & $-0.04$ \\
\bottomrule
\bottomrule
\end{tabular}
\label{tab:harmonics 4 blocks}
\end{table}

\begin{figure}[htbp]
\includegraphics[width=0.45\textwidth]{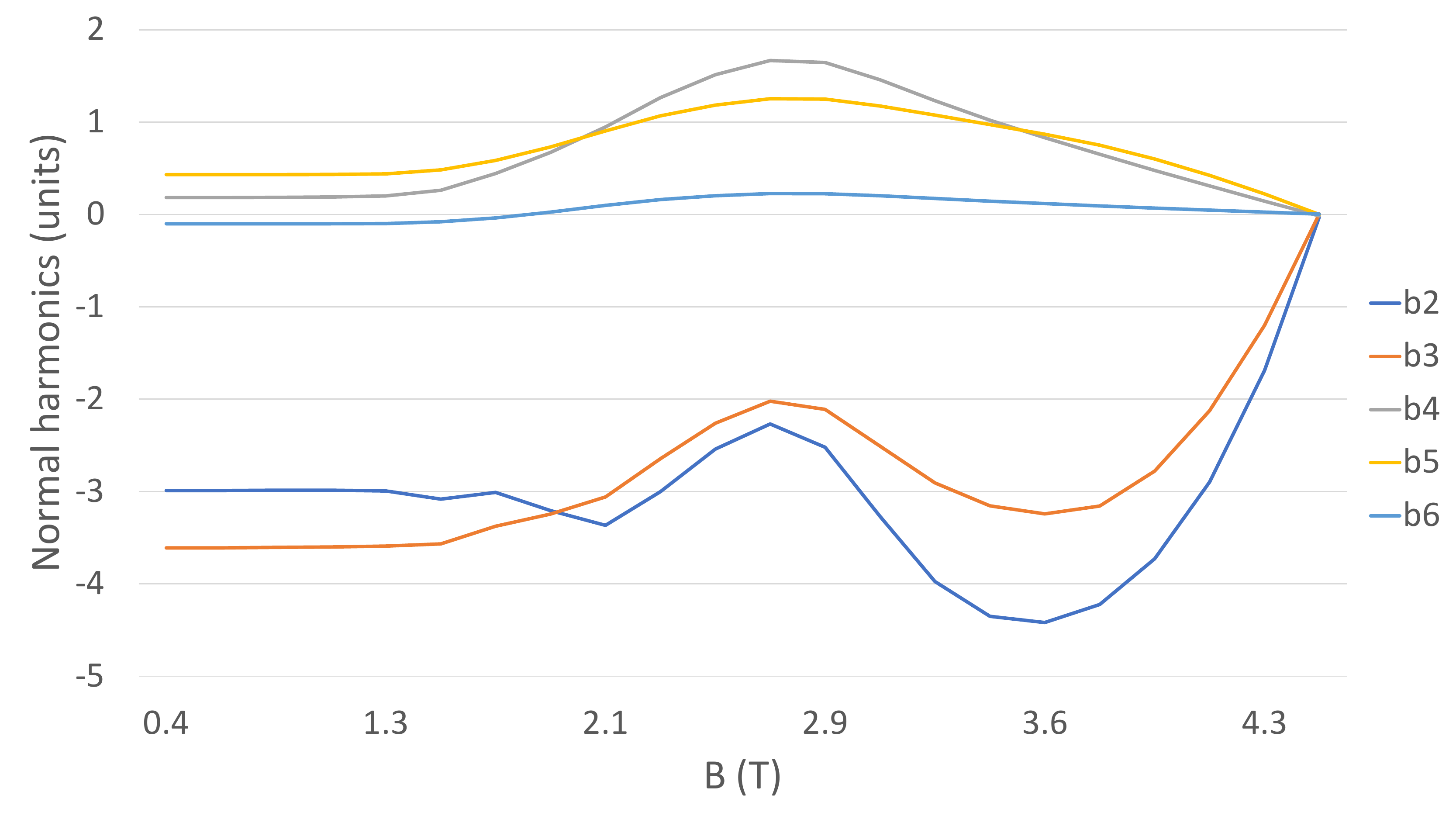}
\caption{Geometrical and saturation normal harmonics from $b_2$ to $b_6$ versus the magnetic field in the bore for the 5 blocks configuration.}
\label{fig:transient 5 blocks}
\end{figure}

\begin{figure}[htbp]
\includegraphics[width=0.45\textwidth]{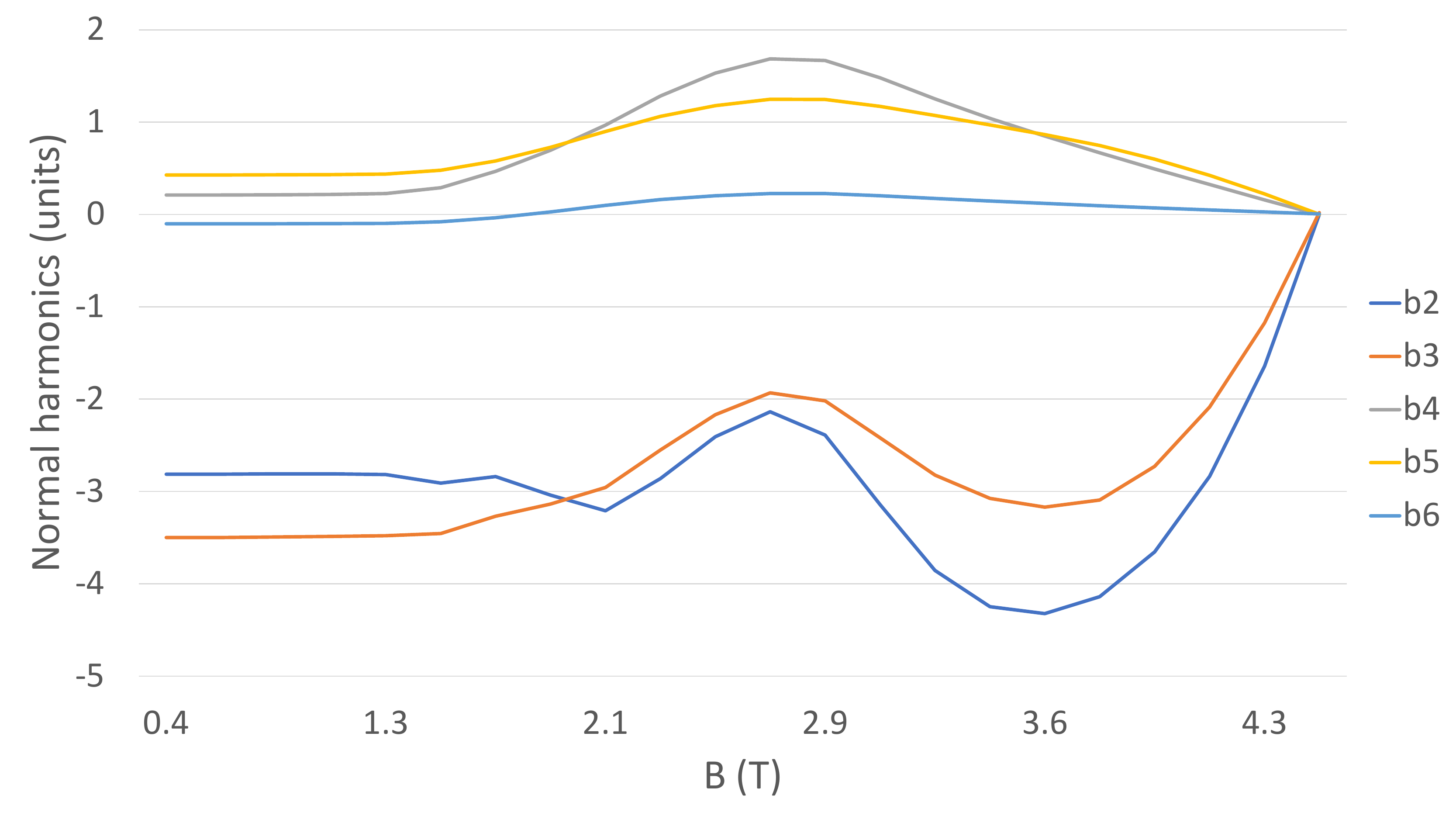}
\caption{Geometrical and saturation normal harmonics from $b_2$ to $b_6$ versus the magnetic field in the bore for the 4 blocks configuration.}
\label{fig:transient 4 blocks}
\end{figure}

The equation systems~\eqref{eq:equation system} and~\eqref{eq:equation system 2} have been solved using the software Wolfram Mathematica 11.2~\cite{Wolfram:Mathematica}. The convergence has been very fast (few minutes). The final numerical optimization with real blocks and iron yoke has been performed by the program ROXIE~\cite{Russenschuck:field computation}, assuming the iron yoke as in Fig.~\ref{fig:D2}. We found two possible electromagnetic designs for the D2 dipole. Fig.~\ref{fig:D2 5 blocks} and~\ref{fig:D2 4 blocks} show the two designs with $5$ and $4$ blocks, respectively. Table~\ref{tab:harmonics 5 blocks} and~\ref{tab:harmonics 4 blocks} display the field quality in the bore for the two configurations, respectively. In the first, the current intensity in each block is $12.8$~\si{\kilo\ampere}, the peak field is $5.32$~\si{\tesla} and the percentage on the load line is about $68.3$\%. In the second, the current intensity in each block is $12.72$~\si{\kilo\ampere}, the peak field is $5.35$~\si{\tesla} and the percentage on the load line is about $68.3$\%. Finally, Fig.~\ref{fig:transient 5 blocks} and Fig~\ref{fig:transient 4 blocks} show the geometrical and saturation normal harmonics from $b_2$ to $b_6$ versus the magnetic field in the bore for both configurations. No possible configuration was found under $4$ blocks.

The five block configuration is not far from the one used in the design of D2, the field quality is slightly better but the peak field is 0.1 T higher. In this case the optimum configuration has been found in more straight way and in much less time. The four block configuration is equivalent to the five block in terms of field quality and in principle is a valid alternative to the present design, which, however is supported now by model construction, whilst the four block option would still require a long development of specific constructive methods.

\section{Conclusions}

Starting from the sector model, we have developed a semi-analytical model for the electromagnetic design of twin aperture $\cos\theta$ superconducting dipoles. It enables to find optimized electromagnetic designs, solving trigonometric equation systems in a short time and this makes possible to map the phase space. As example we showed its application on the D2 dipole~\cite{Farinon:D2} for the High Luminosity upgrade~\cite{CERN:HL-LHC} of LHC~\cite{CERN:LHC}. It allowed to find two possible electromagnetic designs with $4$ and $5$ blocks and with an excellent homogeneity of the magnetic field.

Finally, it is worth noting that the same approach can be used for different coil layouts as for instance dipoles in block coil and common coil configurations~\cite{Rochepault:block coil, Xu:common coil} or quadrupoles.

\end{document}